\documentclass{article}

\usepackage{graphicx}
\usepackage{amssymb}   			
\usepackage{amsthm} 			
\usepackage{amsmath}			
\usepackage{bm}				
\usepackage{savesym}			
\savesymbol{corresponds}		
\usepackage{mathabx}  			
\restoresymbol{ABX}{corresponds}	

\usepackage{hyperref}
\usepackage{float}

\title{A general local causality principle of space-time}

\author{
Benjam\'{\i}n Calvo-Mozo\\
Observatorio Astron\'omico Nacional, Facultad de Ciencias\\ Universidad Nacional de Colombia\\ Carrera 30 \# 45-03, edificio 413, 111321 Bogot\'a D.C.,  Colombia\\ 
bcalvom@unal.edu.co}

\begin{document}
\maketitle

\begin{abstract}
We propose a generalisation of the local causality principle of space-time, asserting that it holds for all regimes of motion, including superluminal motions. It assumes the existence of a countably infinite set of metrical null cone speeds, $c_k$, where the first one, $c_1=c$, corresponds to the speed of light in vacuum. Our associated space-time measures do not diverge at the maximum speed of each interval of speeds and implies a generalisation of Einstein's rule for velocities addition. We construct a causal structure for each regime of motion. After introducing a simple dynamical measure, we derive an expression for the energy of material particles, which approaches the relativistic one when $v<c$. An experiment to energise photons in an 1-1 process is proposed as a test of our interpretation of the non divergence at the speed of light of present space-time measures. We discuss also the possible transition of a material particle from the subluminal regime $v<c$ to the first superluminal regime and vice versa, making discrete changes in $v^2/c^2$ around the unit in terms of a very tiny constant, $\epsilon^2$, which we introduce to prevent the divergence of the Lorentz $\gamma$ factor at the speed of light in vacuum. 
\end{abstract}

\maketitle

\section{Introduction}
The local causality principle of space-time is one basic pillar of the general theory of relativity\cite{Einstein1916}, and as such is the first out of three postulates on which Hawking and Ellis \cite{Hawking-Ellis} developed their space-time physics. They exclude tachyons from their formal description, for this kind of particles follow spacelike intervals instead of timelike intervals associated to known material particles. Despite this difficulty, various researchers worked on superluminal particles or tachyons. For instance, Hill and Cox \cite{Hill-Cox2012} used the current Lorentz transformations for subluminal particles while for superluminal particles they changed the Lorentz $\gamma$ factor for $(v^2/c^2-1)^{-1/2}$, a real number for $v>c$. Further, they partitioned speed values for material particles in two ranges, $0\leq v< c$, and the other with superluminal motions, $c<v<\infty$, that is, in their work superluminal speeds are not upper bounded. The existence of an upper speed $c$ is fundamental to construct a causal structure on space-time for the subluminal regime of motions.

Present work proposes instead, a generalisation of the local causality principle of space-time for all regimes of motion, the subluminal and superluminal ones, where we call the subluminal regime of motion that for which we obtain the interval on the reals $0\leq v/c\leq 1$. Einstein's rule for velocities addition \cite{Einstein1905} warrants this inequality -cf. eq.(\ref{sumaE}). The generalisation of the local causality principle of space-time that we propose, is related to the existence of a countably infinite set of metrical null cone speeds $c_k$, with $k\geq1$ a natural number, such that $ c_{k+1}\gg c_k $, and $ c_1=c $, the speed of light in vacuum. Then, if we allow that $v/c$ runs over all non-negative real numbers, this set is partitioned in pairwise disjoint sets, whose union gives the complete set. The first set is closed, that associated to the subluminal regime of motion, while the others are half-closed, at the top value $c_k$, $k\geq2$, which correspond to superluminal regimes of motion. Therefore, a generalisation of Einstein's rule for the addition of velocities should hold for all regimes of motion, such that $c_k$ remains the top speed for every regime of motion, i.e. for all $k\geq1$. 

We find then the associated space-time measures which are compatible with the generalised rule for velocities addition. We begin by exploring a generalisation of Lorentz transformations, which keeps invariant the speed of light in vacuum and also become finite at this limit speed. We carry it out in two steps. First, we add a very tiny dimensionless constant, say, $\epsilon^2$, within the square root of Lorentz transformations, in order that its inverse becomes finite at the speed of light value; let $\gamma_1$ be the non-divergent term. As a second step, we change the equations for $y$ and $z$, because one requires that the new set of space and time measures keeps invariant the speed of light in vacuum. The new expressions for $y$ and $z$ are such that $y/y'$ and $z/z'$ equal $\beta_1\gamma_1$, where $\beta_1$ is the old radical, this time in the numerator. One can write down the set of space-time equations in a matrix form, namely, as the factor $\beta_1\gamma_1$ times an inverse Lorentz matrix; therefore, the new set of space-time measures can be described as obtained from a Lorentz transformation followed by a regularisation. Einstein's rule for velocities addition remains the same. Space-time measures of present work approaches Lorentz transformations for most relativistic speeds, which can be written as the condition $\beta_1^2\gg\epsilon^2$. For the speed of light we have only one degree of freedom: $y=z=0$ and $x=ct=\epsilon^{-1}(x'+ct')$; further, we also obtain: $x'=ct'=\epsilon x/2$.

Space-time measures for superluminal regimes of motion are obtained using factors $\beta_k$, $\gamma_k$, $k\geq2$, that is, keeping the same algebraic form of the space-time measures in the subluminal regime of motion, and changing the subindex $1$ by the subindex $k$. We impose the condition that the factor $\beta_k$ vanishes while $\gamma_k=k\epsilon^{-1}$ at the top speed of every regime of motion. The form of all $\gamma_k$ factors are obtained using the previous condition of its value at the respective top speed, and then we find that $c_k/c\simeq\epsilon^{-k+1}$  for all $k\geq2$. The inverse measures are obtained through a matrix method. We find the values of lengths of moving rods and of time intervals of clocks in motion.

Minkowski \cite{Minkowski} introduced space-time as a metric space, in which intervals are invariant, being timelike intervals for material particles while light has associated null intervals. In present work, we define first a k-quadratic form for every pair of events, and later the so called ``k-interval", taking into account $c_k^2 dt^2$ and $dx^2$, $dy^2$, $dz^2$. Thus, k-intervals are k-timelike or k-null, depending if they are positive or zero, respectively. We define also k-timelike and k-null curves. Given some space-time event, say $p$, we define the region of space-time called here the k-chronological future of $p$, as the region whose events are linked from $p$ through future oriented k-timelike curves. There is also a region of space-time composed by events which can be joined from $p$ through future oriented k-null paths, denoted here $N_k^+(p)$. The union of the two regions gives the set of events $J_k^+(p)$, called here the k-causal future of $p$. Similarly, it is possible to construct the sets of events which arrive to $p$ following the arrow of time, through either k-timelike or k-null paths, to obtain the sets $I_k^-(p)$, $N_k^-(p)$, the k-chronological past of $p$, and past k-null cone, respectively, and their union gives the set $J_k^-(p)$, called here the k-causal past of $p$. The achronal hypersurface associated to an event $p$, is defined as the set of all events with a negative value of the previously mentioned k-quadratic form with respect to $p$, for all integers $k\geq1$; we add the event $p$, too.

We make an interpretation of the finite and non divergent form which acquire the space-time measures of present work at the speed of light $c=c_1$. These measures refer to points, though the finest partition on space is of order $L_P^3$, where $L_P$ is Planck's length; we call here an {\it element} of space to this finest partition. Carath\'eodory's measure theory \cite{Caratheodory} defined on boolean rings of {\it somas} is suitable for our description. A structure moving with the speed of light is interpret then as some kind of structure such that half of their elements are occupied while the other half will be occupied a time $\lambda/(2c)$ later. The situation replicates itself after an elapsed time $\lambda/c$ along the direction of propagation a distance $\lambda$ ahead. With this interpretation in mind, one describes every element of such structure through a phase, and from it one derives generalisations for both, the Doppler effect and the aberration of light, for all structures moving with a metrical null cone speed. We introduce a simple dynamical measure, which at every element of space enables one to derive an expression for the energy of material particles. For occupied elements of structures moving with the speed of light we associate a dynamical measure equals to $h$, while for material particles it equals $h/2$ per occupied element, in the subluminal regime of motion. To test the interpretation we referred to above, a heuristic experiment to energise (or de-energise) photons is proposed as a function of an applied electric potential difference  (i.e. voltage) along photon's path. We give the expression for it, and ours is a one to one photon's energy conversion. An estimated value for the new tiny constant as $\epsilon^2\simeq3.72\times10^{-54}$ is inferred. In the final discussion one examines the possibility of a discrete transition around the speed of light value, in both senses, for a material particle, such that $v^2/c^2$ changes around the unit in terms of $\epsilon^2$. The previously mentioned dynamical measure plays an important role to carry out this discrete transition.

\section{Generalising local causality}
If one defines a local light cone at every event of space-time then one can relate chronologically all events at the region of space-time within this light cone with respect to the former event, that is, one can trace a timelike path between the given event to any other event within the associated light cone. If we extend the region to include also the light cone as well as its interior, then every event of the new extended region can be joined with either a timelike or a null future oriented path to/from the given event on which one defines the light cone. In a flat space-time we need a Minkowski metric to carry out a causal structure on these bases; it is compatible with Lorentz transformations, which implies the invariance of the speed of light in vacuum, thus, it becomes a maximum speed. Ponderable matter particles always follow timelike curves in space-time and then have speeds less than the speed of light ($0\leq v<c$) in vacuum. Metrics conformal with the Minkowki metric preserve the causal structure.

Let us assume that there exists the possibility to have material particles with speeds greater than the speed of light in vacuum, called tachyons. In author's opinion, if tachyons exist, they should show an ordering of events in the sense of following generalised timelike paths in space-time, for nature shows us that events occur in a time ordered sequence. It implies the existence of at least a second speed, say $c_2\gg c$, which serves as the top and invariant speed for the first range of superluminal motions. Thanks to $c_2$, we can think of a second Minkowski-like metric, valid for material particles in superluminal motions with speeds in the range $c<v<c_2$. If we accept superluminal motions and consider all speeds greater than the speed of light in vacuum, then there should exist also many invariant and top speeds for superluminal motions in intervals, say, $c_3\gg c_2$, $c_4\gg c_3$, and so on, in order to maintain local causality if these special speeds are associated to superluminal metrical null cones. In other words, we can generalise the local causality principle, which we state as follows:\\

{\em Space-time is locally causal at all regimes of motion, which are determined by a countably infinite set of speeds associated to metrical null cones, being the speed of light in vacuum the first of them}.\\

In this way, motions are divided into intervals of speed, called here regimes of motion, of which the first one, referred to as the subluminal regime in present work, corresponds to the closed interval of real numbers $0\leq v/c\leq1$, while the other regimes of motion have associated semi-open or half-closed intervals on the reals, right-closed, that is, the end point is the real number associated to the dimensionless speed of the corresponding null cone speed, $c_{k-1}/c<v/c\leq c_{k}/c$, for all superluminal regimes, $k\geq2$. The interval of dimensionless speeds  associated to the first superluminal regime is $1<v/c\leq c_2/c$. Later in this work, we find a rule for computing the $c_k$ speeds associated to superluminal metrical null cones. Thus, dimensionless speeds $v/c$ run over all nonnegative real numbers and each value of it belongs to one of the pairwise disjoint sets, whose union equals the set of all dimensionless speeds. In set theory, if one has a nonempty set, say the set of dimensionless speeds assuming both subluminal and superluminal motions, the axiom of choice \cite{Zermelo1904} stipulates that if we decompose the original nonempty set in pairwise disjoint and nonempty sets (a class of sets) whose union gives the former set, then there exists another set whose elements are given by selected elements, one per every set of the class of sets. In present case, the selected set is the set of the dimensionless speeds of the metrical null cones of space-time, that is, $C=\{c_1/c, c_2/c, c_3/c, \ldots\}$, where $c_1/c=1$.

\section{Space-time measures}

Lorentz transformations can be viewed as space and time measures which implies the invariance of the speed of light in vacuum, c, for inertial observers in uniform relative motion. Einstein's addition of velocities warrants the invariance of the speed of light, and operatively takes $c$ also as the maximum speed for the propagation of information and interactions. If for the sake of simplicity, we consider motion only along a direction, we can express velocities addition as:
\begin{equation}\label{sumaE}
u\oplus w=\frac{u+w}{1+uw},
\end{equation}

\noindent where $u=v/c$ denotes the dimensionless speed, with respect to the speed of light in vacuum, of the relative speed between two inertial frames, and $w=V/c$ stands for the dimensionless speed of something relative to one of the frames. Algebraically, we can take in eq.(\ref{sumaE}) $u$ or $w$ as the unit and obtains $u\oplus w=1$ in both cases. We can generalise eq.(\ref{sumaE}) in order that $c_k$, $k\geq 1$ becomes the maximum speed of the respective regime of motion, either in the subluminal or in a superluminal regime. We can do it if instead of $u$, $w$ we use in this equation $u_k=v/c_k$, $w_k=V/c_k$.

Lorentz transformations do not allow the case $u=1$ due to the divergence of them in that case, which is compatible with the relativistic interpretation that there are not inertial frames moving with the speed of light in vacuum. Let us consider space and time measures which preserve the invariance of the speed of light in vacuum and which do not diverge at this top speed value; however, the case $v=c$ deserves an interpretation as we shall do later. We introduce the non divergence at the speed of light in vacuum of our space and time measures to assure that both, these measures and the rule for velocities addition holds for the complete interval of speeds of this regime of motion, the subluminal one. Further, it gives us a way to calculate the top speed for every superluminal regime. Let $\epsilon^2$ be a real positive dimensionless constant, very small with respect to the unit; later in this paper we will estimate its value as of the order of $10^{-54}$ -cf. eq.(\ref{epsilon2}). Let us add this constant $\epsilon^2$ within the square root which appears in Lorentz transformations, such that the new non divergent $x$ and $t$ measures are:
\begin{equation}\label{Leps-1}
x=\frac{x'+uct'}{\sqrt{1+\epsilon^2-u^2}},\quad ct=\frac{ct'+ux'}{\sqrt{1+\epsilon^2-u^2}},
\end{equation}
where $u=v/c$. To preserve the speed of light in vacuum as the maximum and invariant speed for all inertial frames, we ought to modify the expressions for $y$ and $z$ as well:
\begin{equation}\label{Leps-2}
y=\left(\frac{1-u^2}{1+\epsilon^2-u^2}\right)^{1/2} y',\,\, z=\left(\frac{1-u^2}{1+\epsilon^2-u^2}\right)^{1/2} z'.\\
\end{equation}
We see that eqs.(\ref{Leps-1}),(\ref{Leps-2}) approach Lorentz transformations for most relativistic speeds, that is, when the condition $\beta_1^2=(1-v^2/c^2)\gg \epsilon^2$ holds. We can rewrite eqs.(\ref{Leps-1}),(\ref{Leps-2}) in a different algebraic form, which enables their generalisation to other regimes of motion. First, let $\gamma_1$ be the non divergent term (at the speed of light) in these expressions:

\begin{equation}\label{gamma1}
\gamma_1 = (1+\epsilon^2-u^2)^{-1/2},\quad u=v/c.
\end{equation}

This gamma factor equals $\epsilon^{-1}$ when $v=c$, which is a large quantity but anyway a finite one. With the gamma factor $\gamma_1$, our $x$, $t$ measures given by eqs.(\ref{Leps-1}) are:

\begin{equation}\label{ecsxt1}
x = \gamma_1(x'+u_1c_1t'),\quad c_1t =\gamma_1(c_1t'+u_1x'),
\end{equation}

\noindent where $c_1=c$, $u_1=v/c_1$, and the $\gamma_1$ factor is given by eq.(\ref{gamma1}). Now, our expressions for the $y$ and $z$ measures given by eqs.(\ref{Leps-2}) in the new algebraic form are:

\begin{equation}\label{ecsyz1}
y = \beta_1\gamma_1 y',\quad z = \beta_1\gamma_1 z',\quad \beta_1=(1-u_1^2)^{1/2},
\end{equation}

We see that for the speed of light in vacuum, $v=c$, eqs.(\ref{ecsyz1}) imply the vanishing of the $y,z$ measures, while at this speed eqs.(\ref{gamma1}),(\ref{ecsxt1}) give,

\begin{equation}\label{luz_1}
x = ct= \epsilon^{-1}(x'+ct').
\end{equation}

That is, something moving with the speed of light kinematically reduces its description to one degree of spatial freedom. From eqs.(\ref{ecsxt1}),(\ref{ecsyz1}) we can obtain Einstein's rule for velocities addition, which warrants the invariance of the speed of light in vacuum for inertial observers in uniform relative motion; it also implies that $c$ becomes the maximum speed at the subluminal regime of motions. Our new space-time measures as given by eqs.(\ref{ecsxt1}),(\ref{ecsyz1}) can be seen as a Lorentz transformation followed by a regularisation transformation, which is stipulated through the multiplication of the (inverse) Lorentz transformation expressions by the factor $\beta_1\gamma_1$.

We can generalise the space-time measures for any regime of motion $k\geq 1$, if instead of the subindex 1 given above, we use the index $k$ keeping in mind that: 
\begin{equation}\label{conditions}
\gamma_k(c_k)=k\epsilon^{-1}, \beta_k(c_k)=0.
\end{equation}

We do it because a generalisation of eqs.(\ref{ecsxt1}),(\ref{ecsyz1}) should reduce to them for the subluminal regime, i.e. for $k=1$. In this way, our new and general space-time measures are:

\begin{equation}\label{ecsxtk}
x = \gamma_k(x'+u_kc_kt'),\quad c_kt =\gamma_k(c_kt'+u_kx'),
\end{equation}

\noindent for the $x$ and $t$ measures, while the generalisation of eqs.(\ref{ecsyz1}) are:
\begin{equation}\label{ecsyzk}
y = \beta_k\gamma_k y',\quad z = \beta_k\gamma_k z',\quad \beta_k=(1-u_k^2)^{1/2},
\end{equation}

\noindent where $u_k=v/c_k$. We can easily check that eqs.(\ref{ecsxtk}),(\ref{ecsyzk}) imply a generalisation of the rule for velocities addition in dimensionless variables:
\begin{equation}\label{speedsk}
U_{k,x}=\frac{U'_{k,x'}+u_k}{1+u_kU'_{k,x'}},\,\,U_{k,y}=\frac{U'_{k,y'}\beta_k}{1+u_kU'_{k,x'}},\,\,U_{k,z}=\frac{U'_{k,z'}\beta_k}{1+u_kU'_{k,x'}},
\end{equation}
\noindent where ${\bf U}_k={\bf V}/c_k$, ${\bf U'}_k={\bf V'}/c_k$, and ${\bf V}$, ${\bf V}'$ are the velocities of some material particle in some regime of motion $k$ as seen by two inertial observers, say $\Sigma$, $\Sigma'$, respectively, and frame $\Sigma'$ moves with speed $v=u_kc_k$ (velocity ${\bf v}$) with respect to frame $\Sigma$ along the positive x-axis. The Cartesian components of ${\bf U}_k$ are written as $U_{k,x}$,$U_{k,y}$,$U_{k,z}$; we denote similarly the respective components of ${\bf U}'_k$, as the primed variables $U'_{k,x'}$, $U'_{k,y'}$, $U'_{k,z'}$.

Eqs.(\ref{ecsxtk}),(\ref{ecsyzk}) can be written in a matrix form as:
\begin{equation}\label{Lk}
X=\beta_k\gamma_k\Lambda'_k X'=L_k X', \quad L_k=\beta_k\gamma_k\Lambda'_k,
\end{equation}
\noindent where $\Lambda'_k$ can be seen as a generalised (inverse) Lorentz matrix:
\begin{equation}\label{Lmatrix}
\Lambda_k'=
\begin{bmatrix}
\beta_k^{-1} & u_k\beta_k^{-1} & 0 & 0 \\
u_k\beta_k^{-1} & \beta_k^{-1} & 0 & 0 \\
0 & 0 & 1 & 0 \\
0 & 0 & 0 & 1
\end{bmatrix}.
\end{equation}

Let us observe that in the subluminal regime of motion ($k=1$) our matrix $\Lambda'_1$ equals exactly the well known (inverse) Lorentz matrix. We can also write eqs.(\ref{ecsxtk}),(\ref{ecsyzk}) in a 3-vector form as:
\begin{equation}
\begin{gathered}\label{vector-form}
{\bf r}= \beta_k\gamma_k{\bf r}' + \gamma_k\Bigl[(1-\beta_k)\frac{{\bf u}_k\cdot {\bf r}'}{u_k^2}+c_k t' \Bigr]{\bf u}_k,\\
c_k t= \gamma_k\left( c_k t' +{\bf u}_k\cdot{\bf r}' \right).
\end{gathered}
\end{equation}
Then, we have that for $v=c_k$, for all $k\geq1$, there is only  one degree of spatial freedom:
\begin{equation}\label{vector-c_k}
{\bf r}=c_k t\, \hat{{\bf e}}=k\epsilon^{-1}\left({\bf r}'\cdot\hat{{\bf e}} + c_k t'\right)\hat{{\bf e}},
\end{equation}

\noindent where $\hat{{\bf e}}={\bf u}_k/u_k$ denotes the direction of propagation.

The inverse matrix of $\Lambda_k'$, i.e. $\Lambda_k$, is obtained from eq.(\ref{Lmatrix}) changing the sign of $u_k$. Thus, the inverse of eqs.(\ref{ecsxtk})-(\ref{ecsyzk}) can be obtained through $X'=\beta_k^{-1}\gamma_k^{-1}\Lambda_k X$ -cf. eqs.(\ref{Lk}):

\begin{align}\label{inversask}
c_kt'= &\,\beta_k^{-2}\gamma_k^{-1}(c_kt-u_kx),\\ 
x'= &\,\beta_k^{-2}\gamma_k^{-1}(x-u_kc_kt),\label{inversask1}\\ 
y'= &\,\beta_k^{-1}\gamma_k^{-1}y,\,\, z'=\beta_k^{-1}\gamma_k^{-1}z.\label{inversask2}
\end{align}

From these equations we can obtain a rule for speeds addition similar in form to eqs.(\ref{speedsk}), interchanging the components of $\bf U_k$ and $\bf U'_k$, and changing the sign of $u_k$; if after that, we replace $-u_k U_{k,x}$ by ${-\bf u_{k}}\cdot{\bf U_{k}}$, then we can write them in a compact form as:
\begin{equation}\label{speeds-inv}
U'^{2}_{k}=1-\frac{(1-U^{2}_{k})(1-u^{2}_{k})}{(1-{\bf u_{k}}\cdot{\bf U_{k}})^{2}},
\end{equation}

Lengths of moving bodies can be deduced by simultaneously measuring their extreme points. Therefore, from eqs.(\ref{inversask1}),(\ref{inversask2}) one obtains the lengths of moving bodies or structures for any regime of motion labelled by $k\geq1$: 
\begin{equation}\label{lengthsk}
l_{\parallel}(u)=\beta_k^2\gamma_k l_o,\,\,  l_{\perp}(u)=\beta_k\gamma_k l_o,
\end{equation}
\noindent where the first refers to lengths parallel to the body's motion and the second describes lengths when looking at a direction perpendicular to the motion. For moving identical clocks, we compare their time marks using the second of eqs.(\ref{ecsxtk}), and obtain (using $\Delta x'=0$):
\begin{equation}\label{timek}
\Delta t(u)=\gamma_k\Delta t_o.
\end{equation}

The third of eqs.(\ref{ecsyzk}) gives the factor $\beta_k$, valid for all regimes of motion; it vanishes at the corresponding top speed, which is the speed associated to the metrical null cone of the respective regime of motion. We will find now the appropriate expressions for the $\gamma_k$ factor for all superluminal regimes of motion; eq.(\ref{gamma1}) gives us the subluminal $\gamma_1$ factor, which equals $\epsilon^{-1}$ for $v=c$. To find an expression for $\gamma_2$, let us take into account that it should increase with speed as does the $\gamma_1$ factor, and $\gamma_2$ comes with an accumulated $\epsilon^{-1}$ from the top value of the previous regime of motion, the subluminal regime. Further $\gamma_2=2\epsilon^{-1}$ when $v=c_2$, in agreement with the condition given by the first of  eqs.(\ref{conditions}). Thus, we can obtain the $\gamma_2$ factor combining positive and negative powers of $\epsilon^2$, valid for all $c<v\leq c_2$:

\begin{equation}\label{gamma2}
\gamma_2 =\epsilon^{-1}+(\epsilon^{-2}+1+\epsilon^2-u^2)^{-1/2}, u=v/c.
\end{equation}

From this expression we obtain the new limit speed for the first superluminal regime, $c_2$:

\begin{equation}\label{c_2}
c_2=(\epsilon^{-2}+1)^{1/2}c\simeq\epsilon^{-1}c \simeq5\times10^{26}c.
\end{equation}

This is a huge speed. The value of $c_2$ given above takes into account the estimative value of $\epsilon^2$ as given by eq.(\ref{epsilon2}). In our description, the $\gamma_1$ factor given by eq.(\ref{gamma1}) is valid for squared dimensionless speeds in the range $0\leq u^2\leq 1$, that is, in the subluminal regime, while the $\gamma_2\,$factor given by eq.(\ref{gamma2}) applies in the range of squared dimensionless speeds $1< u^2\leq\epsilon^{-2}+1$, which specifies the first superluminal regime; in both cases $u=v/c$. At the second limit speed, $c_2$, eqs.(\ref{ecsxtk}),(\ref{gamma2}), give $x=c_2t=2\epsilon^{-1}(x'+c_2t')$, which is similar in form to eq.(\ref{luz_1}) except for a factor of 2, whilst measures $y,z$ vanish.

We will find now the appropriate expressions for $\gamma_k$, $c_k$ in the general superluminal case; the one for $\beta_k$ is given by eq.(\ref{ecsyzk}). We obtain $\gamma_k$ for any $k\geq2$, considering that it has accumulated $(k-1)\epsilon^{-1}$ from the previous speeds range and that it should equal to $k\epsilon^{-1}$ at the top speed $c_k$ of the respective speeds interval. Further, as in the expression for $\gamma_2$ there are terms with positive and negative powers of $\epsilon^2$, then we shall use higher positive and negative powers of it keeping symmetry in these powers, that is, if there appears the power $\epsilon^{-2j}$, then there appears also the power $\epsilon^{2j}$. And as the first superluminal regime is one of these cases, the operative form of $\gamma_k$ should reduce for $k=2$ to the expression of $\gamma_2$ given by eq.(\ref{gamma2}). In this manner we have:

\begin{equation}\label{gammai}
\gamma_k =(k-1)\epsilon^{-1}+ \epsilon^{k-2}\left[\sum_{j=1}^{k-1}\left(\epsilon^{-2j}+\epsilon^{2j}\right)+1-u^2\right]^{-1/2},\,\, k\geq2, u=v/c.
\end{equation}

From this expression and the condition imposed to the factor $\gamma_k$ at the maximum speed of the associated range stipulated by the first of eqs.(\ref{conditions}), we obtain the expression for $c_k$, $k\geq2$:

\begin{equation}\label{luz-i}
(c_k/c)^2 =\sum_{j=1}^{k-1}\epsilon^{-2j} + \sum_{j=0}^{k-2}\epsilon^{2j},
\end{equation}

\noindent which has the expression given by eq.(\ref{c_2}) as a particular case. For any $k\geq2$, eqs.(\ref{c_2}),(\ref{luz-i}) tell us that in first approximation $c_k/c\simeq\epsilon^{-k+1}\gg1$.  For all $v=c_k$ one has that $y=z=0$, and:

\begin{equation}\label{xt-luz-i}
x=c_kt=k\epsilon^{-1}(x'+c_kt'),\quad \text{and} \quad x'=c_k t'=\epsilon x/(2k)=\epsilon c_kt/(2k),
\end{equation}

\noindent which is a generalization of eq.(\ref{luz_1}). The first of eqs.(\ref{xt-luz-i}) is obtained directly from eqs.(\ref{conditions}),(\ref{ecsxtk}), making $u_k=1$, that is, for $v=c_k$. The second of eqs.(\ref{xt-luz-i}) is obtained through a limit procedure applied to eqs.(\ref{inversask}),(\ref{inversask1}), which uses the L'H\^opital rule of calculus.\\
Let us observe that if we take into account eq.(\ref{luz-i}), the expression for $\gamma_k$ given by eq.(\ref{gammai}) can be re-written as:
\begin{equation}
\gamma_k =(k-1)\epsilon^{-1}+ \epsilon^{k-2}\left[(c_k/c)^2+\epsilon^{2(k-1)}-v^2/c^2\right]^{-1/2},\,k\geq2,
\end{equation}

\section{Causal structure}
Let $\mathcal{M}$ be the set of all events in space-time, a 4-dimensional manifold, such that each point or event $p$ can be written in components as $(x^0,x^1,x^2,x^3)$, of which the former contains time, $x^0=c t$. We can define on $\mathcal{M}$ a quadratic form generalising the one proposed by Zeeman \cite{Zeeman1964} and by Kronheimer and Penrose \cite{Kronheimer-Penrose1967}, such that if $p$ and $q$ are events, with components $x^{\alpha}$, $y^{\beta}$, respectively, with $\alpha,\beta:0,1,2,3$, then:
\begin{equation}\label{2-form}
Q_k(p,q)= (c_k/c)^2(x^0 - y^0)^2 - \sum_{i=1}^{3}(x^i - y^i)^2.
\end{equation}

If the event $q$ lies in some small open neighborhood of the event $p$, there exists the possibility to have a differential which can be called a "k-interval":
\begin{equation}\label{k-interval}
ds^2_k=c^2_k dt^2 - dx^2 - dy^2 - dz^2,
\end{equation}
\noindent where $c_k dt$, $dx$, $dy$, and $dz$ can be obtained from eqs.(\ref{ecsxtk}),(\ref{ecsyzk}) with $u_k$, $\beta_k$, $\gamma_k$ constants; $c_k$ is given by eq.(\ref{luz-i}) for $k\geq2$ and $c_1=c$. For each $k\geq1$, $k$ an integer, one has a regime of motion, that is, an interval of speeds for which there is a top and invariant speed $c_k$ and a set of space-time measures given by eqs.(\ref{ecsxtk}),(\ref{ecsyzk}). We can say that k-intervals are k-timelike or k-null depending on the value of $ds^2_k$ as given by eq.(\ref{k-interval}), namely, if it is positive or zero, respectively. From eq.(\ref{k-interval}) one obtains a metric ${\bf g}_k$ for every regime of motion, which results to be conformal to a Minkowskian metric $\mbox{\boldmath $\eta$}$ for the primed variables:
\begin{equation}\label{k-metric}
{\bf g}_k=\beta_k^2\gamma_k^2\mbox{\boldmath $\eta$}.
\end{equation}

The notions of causal precedence ($\prec$) and of chronological ($\lll$) precedence as developed by Kronheimer and Penrose \cite{Kronheimer-Penrose1967}, can be generalized for all regimes of motion distinguishing them with an index $k\geq1$, taking into account the causality and chronological relations as described by Carter \cite{Carter1971}, such that given two events of the space-time manifold, say, $p,q\in \mathcal{M}$, then $p$ causally precedes $q$ with respect to some "auxiliary set" $\mathcal{U}\subset \mathcal{M}$,

\begin{equation}\label{causal-k}
p\underset{\mathcal{U}}{\prec} q,\, \text{if}:\, x^0<y^0\,\, \text{and}\,\, Q_k(p,q)\geq0,
\end{equation}

\noindent under the condition to be restricted to the subset $\mathcal{U}\subset\mathcal{M}$. For subluminal motions, $k=1$, and our set $\mathcal{U}$ is confined to the light cone and its interior; let us denote it as $\mathcal{U}_1$. For $k=2$, the first superluminal regime, the auxiliary set $\mathcal{U}$, say, $\mathcal{U}_2$, is restricted to the region within the light cone (excluding it) and the second null cone, including it. In general, for any regime of motion distinguished by some $k\geq2$, the corresponding set $\mathcal{U}_k$ is restricted to the region between the ($k-1$)-null cone, excluding it, and the $k$-null cone, taking it as part of this auxiliary set. Then, taking into account all regimes of motion, labelled by an integer $k\geq1$, we have the following partial orderings in space-time for all cases:

\begin{equation}\label{k-relations}
\begin{split}
p\underset{\mathcal{U}_k}{\prec} q, &\quad \text{if}:\, x^0\leq y^0\,\, \text{and}\,\, Q_k(p,q)\geq0,\\
p\underset{\mathcal{U}_k}{\lll} q, &\quad \text{if}:\, x^0\leq y^0\,\, \text{and}\,\, Q_k(p,q)>0,\\
p\underset{\mathcal{U}_k}{\to} q, &\quad \text{if}:\, x^0\leq y^0\,\, \text{and}\,\, Q_k(p,q)=0.
\end{split}
\end{equation}

In analogy with the definitions given by Kronheimer and Penrose \cite{Kronheimer-Penrose1967}, we can call the above relations, k-causal precedence, k-chronological precedence and k-horismos, respectively. In the respective subset of space-time associated to some pair of events in the same regime of motion, in which one of them causally or chronologically precedes the other, one can link the two through a continuous succession of events, that is, we link them by a simple curve in space-time. Any curve on the space-time manifold is conceived as usual, that is, as a map of an interval of the reals on the space-time manifold; if the curve does not intersect with itself, then it is a simple curve. In present work we say that in space-time, a curve is a k-timelike curve if every pair of events on it, say, $p,q\in\lambda_k$, where $\lambda_k$ is the k-timelike curve, are such that $Q_k(p,q)>0$. We say also that a $\lambda_k$ curve is a k-null curve if for every pair of events on it, $p,q\in\lambda_k$, one has that $Q_k(p,q)=0$. If these curves are future directed, one has further that if they go from event $p$ to event $q$, then $x^0<y^0$.\\

Under the previous partial orderings in space-time and given a regime of motion labeled by $k$, we see that if the event $p$ chronologically precedes the event $q$, then there is at least a k-timelike curve joining them from $p$ to $q$ as time runs forwardly. Similarly, if the event $p$ causally precedes the event $q$, then as time goes on, there is either a k-timelike curve or a k-null curve in the sense from $p$ to $q$. Thus, for any event $p\in\mathcal{M}$, we define the following regions of space-time which lead us to the notion of the causal structure of space-time for all regimes of motion:

\begin{equation}\label{k-regions}
\begin{aligned}
I_k^+(p)=& \{q\in\mathcal{M}: p\underset{\mathcal{U}_k}{\lll} q\}, &I_k^-(p)= \{q\in\mathcal{M}: q\underset{\mathcal{U}_k}{\lll} p\},\\ 
J_k^+(p)=& \{q\in\mathcal{M}: p\underset{\mathcal{U}_k}{\prec} q\}, &J_k^-(p)= \{q\in\mathcal{M}: q\underset{\mathcal{U}_k}{\prec} p\},\\
N_k^+(p)=& \{q\in\mathcal{M}: p\underset{\mathcal{U}_k}{\to} q\}, &N_k^-(p)= \{q\in\mathcal{M}: q\underset{\mathcal{U}_k}{\to} p\}.\\
\end{aligned}
\end{equation}

\begin{figure}
\centering
\includegraphics[scale=0.4]{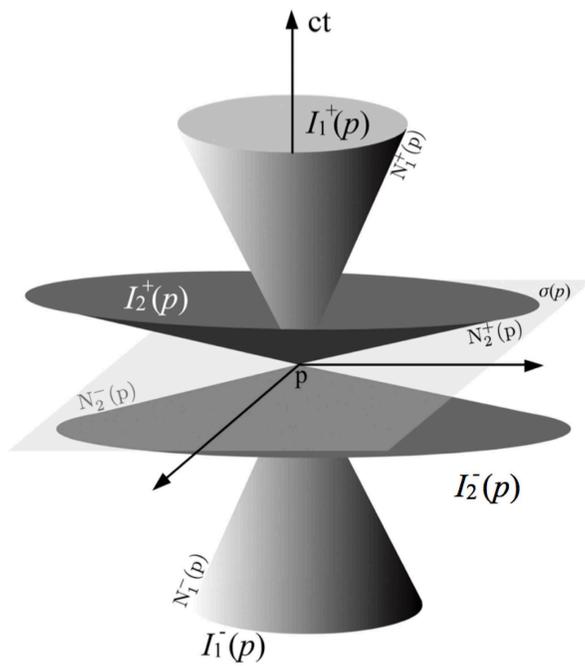}
\caption{Given some event $p$ one has their associated past and future null cones for the subluminal and first superluminal regimes of motion, and the corresponding 1-,2-chronological past and future regions.}\label{null-cones}
\end{figure}

 We call them, the k-chronological future, k-chronological past, k-causal future and k-causal past of the event $p$, respectively, for the first four subsets of space-time associated to $p$, and for the last two above we call them the future and past k-null cones, respectively. In Fig.\ref{null-cones} we represent the first two  future and past metrical null cones associated to some given event $p\in\mathcal{M}$; they are not drawn to scale and are labeled as the $N_1^+(p)$, $N_2^+(p)$ regions in the upper part of the figure and the regions $N_1^-(p)$, $N_2^-(p)$ in the lower part of it. We also see those regions of k-chronological past and future of the event $p$, with $k=1,2$, that is, for the subluminal $I_1^-(p)$, $I_1^+(p)$, and first superluminal $I_2^-(p)$, $I_2^+(p)$ regimes of motion.

In analogy with Kronheimer-Penrose \cite{Kronheimer-Penrose1967} we can say that the set $\mathcal{M}$ together with partial orderings like our k-causal precedence, k-chronological precedence and k-horismos, constitute a causal space; in our case, causality in space-time is satisfied by regions, each one corresponding to a regime of motion, that is, space-time is causally multi-structured. 

We can also have subsets of $\mathcal{M}$, say, $\sigma_p\subset\mathcal{M}$, the elements of which cannot be joined, neither with a $k$-timelike path nor with a $k$-null curve, for all $k\geq1$; these subsets of space-time correspond to achronal hypersurfaces, which can be defined with the aid of eq.(\ref{2-form}) as:

\begin{equation}\label{achronal-s}
\sigma_p= \{ q\in\mathcal{M}:\text{given}\,\, p\in\mathcal{M},\, Q_k(p,q)<0,\,\, \forall k\in\mathbb{N}\}\bigcup\{p\}.
\end{equation}

These spacelike hypersurfaces are 3-dimensional manifolds or {\it slices} of space-time. Let $\sigma_1,\sigma_2$ be two achronal hypersurfaces. Then, if the points (i.e. events) in one of them are reachable from the other through future directed k-timelike or k-null paths, we can label each achronal surface $\sigma$ by some real number $t(\sigma)$, and calls it time, such that $t(\sigma_2)>t(\sigma_1)$ if points on $\sigma_2$ are reachable from some point on $\sigma_1$ through future directed k-null or k-timelike paths. With this procedure we infer a splitting of our space-time, compatible with Geroch's\cite{Geroch1970} splitting. As an example, we show in Fig.\ref{null-cones} a 2-dimensional achronal hypersurface $\sigma_p$, associated to some some event $p$.

\section{An interpretation}
The generalisation of the local causality principle of space-time for all regimes of motion as proposed in present work, implies the existence of a set of top and invariant speeds, metrical null cone speeds, which partition speed values in sets of nonnegative real numbers or intervals, being the first one closed in both sides, that pertaining to the subluminal regime, $0\leq v/c\leq1$, whilst the others are open in the lower end and closed in the upper value, i.e.  $c_{k-1}/c<v/c\leq c_k/c$, $k\geq2$, which correspond to superluminal regimes. We include the top interval speed $c_k$ for all regimes of motion, i.e. for all $k\geq1$, for it is allowed by the rule of velocities composition given by eqs.(\ref{speedsk}), which is a generalisation of Einstein's rule for velocities addition. Eqs.(\ref{ecsxtk}),(\ref{ecsyzk}) satisfy the conditions of non divergence of space-time measures at speed $c_k$ and lead to the above rule of velocities addition. However, something moving with the speed of light $c$ or at other metrical null cone speed $c_k$ do not serve as a frame of reference, for it appears moving at speed $c$, $c_k$ for all inertial observers, respectively, then eqs.(\ref{luz_1}),(\ref{vector-c_k}),(\ref{xt-luz-i}) deserve an interpretation.

When we deal with the theory of relativity in the restricted sense, for subluminal motions, Einstein's measures as seen as marks made through the use of rods and clocks suit well. To interpret both, space and time measures in a more general way, we can opt to make compatible the mathematical and physical points of view of these measures. A first step from the physical point of view is to introduce an appropriate frame of reference for a given observer. From it, one constructs a formal mathematical description of space-time as a 4-dimensional manifold endowed with metric and a causal structure.

From the mathematical side, calculations on manifolds usually employs derivatives and integrals; they require a very fine partition of space-time.  When we go down looking for the finest partition of space, we find lengths on the order of Planck's length, $L_P=(\hbar G/c^3)^{1/2}\simeq1.62\times10^{-35}$m. Planck \cite{Planck1899} thought that when we arrive to such lengths, space could be described in a discrete manner, while for Sakharov\cite{Sakharov1967} lengths of order $L_P$ represent our limits of the concept of space in the sense of localisation. Wheeler proposed pregeometry (see box 44.5 of \cite{Gravitation}),  the logic of propositions, as the basic building reasoning of space-time at Planck's length scale. Smolin \cite{Smolin2001} considers that our finest partition of space is a very tiny volume given by $L_P^3$ and the finest time interval is $t_P=L_P/c$, the Planck time; instead of points he considers tiny volumes or order $L_P^3$. Taking into account these ideas, present work considers that points associated to space-time measures given by eqs.(\ref{ecsxtk}),(\ref{ecsyzk}), or of their inverses given by eqs.(\ref{inversask})-(\ref{inversask2}), are of the order of $L_P^3$ in volume, which we will call here {\it elements} of space. On such elements of space we can define geometrical, kinematical or dynamical measures. 

The measure theory of sets is the basic mathematical theory for analysis. Usually, they are based on $\sigma$-algebras of sets of points. Carath\'eodory \cite{Caratheodory} developed a measure theory of sets of {\it somas} on Boolean rings. Somas, from the Greek word for body, are not points but something extended, which can be very small, like for instance our elements of space, or intervals of real numbers. He introduced the concept of "finitely-valued place functions", which he conceived as a generalisation of point function in Euclidean spaces or other spaces. Inspired in his work we consider that an element positioned by some value $x$ in the first of eqs.(\ref{ecsxtk}) corresponds to some measure function $\mu$ applied to a soma of an appropriate Boolean ring of somas defined on elements of space. Any measure function is such that, if it applies to the void set gives the value zero, and if $A$,$B$ are disjoint somas, then the measure of their union equals the sum of the individual measures. Let us assume that our space and time measures, $x$,$y$,$z$,$c_kt$ are the result of applying three measure functions, say, $\mu_{1k}$,$\mu_{2k}$,$\mu_{0k}$, to somas associated to some element of space $A$, whose position we want to describe at some instant of time, in a given inertial frame of reference centred at element $A_o$, given some regime of motion at element $A$, specified by $k\in\mathbb{N}$. 

Let $B_1$ denote the soma made of elements representing the projection of element $A$ along the $x$-axis, including the projection element and the reference element $A_o$. Further, let $B_2$,$B_3$ be the somas of elements associated to the projection of element $A$ along the $y$-,$z$-axes, respectively, which include the respective projection elements and $A_o$. If we define $L_{\alpha}=B_{\alpha}-A_o$, $\alpha=1,2,3$, where the minus sign denotes the set operation difference, which is the set obtained from the set $B_{\alpha}$ taken away the element $A_o$, then, we have in this description that: 
\begin{equation}\label{medida-yz}
y=\mu_{2k}(L_2)=\beta_k\gamma_k y',\, z=\mu_{2k}(L_3)=\beta_k\gamma_k z',
\end{equation}
If the projection of element $A$ gives a soma containing only the reference element $A_o$, the difference of somas of type $B_{\alpha}-A_o$, equals the void set in that case, and then the respective measure equals zero. For the $x$ and $t$ measures one should take into account also another kind of soma containing intervals of duration $t_P$ each, because the motion takes place along the $x$-axis. Thus, let $\tau$ be the soma containing many times $t_P$ as necessary, as elementary intervals of time, indicating the time mark associated to an event at the element $A$. Its time mark referred to the given reference frame is taken as customary, that is, by a local clock at rest synchronised with the one (identical clock) at frame's origin "point" or element $A_o$. We define then the soma $T=\tau-t_P$, where the elementary time interval $t_P$ corresponds to the event where clocks indicate the mark zero. In this manner, the first of eqs.(\ref{ecsxtk}) can be seen as:
\begin{equation}\label{medida-x}
x=\mu_{1k}(L_1\cup T)=\mu_{1k}(L_1)+\mu_{1k}(T),\, \mu_{1k}(L_1)=\gamma_k x',\, \mu_{1k}(T)=\gamma_k u_kc_kt',
\end{equation}
\noindent while the second of eqs.(\ref{ecsxtk}) can be put in the form:
\begin{equation}\label{medida-t}
c_kt=\mu_{0k}(L_1\cup T)=\mu_{0k}(L_1)+\mu_{0k}(T),\, \mu_{0k}(L_1)=\gamma_k u_kx',\, \mu_{0k}(T)=\gamma_k c_kt'.
\end{equation}

Eqs.(\ref{medida-x}),(\ref{medida-t}) consider somas $L$ and $T$ as disjoint sets because they come from different conceptual bases and our space-time is flat. Carath\'eodory proved a theorem, which states that the measures built on Boolean rings can be extended to measures on the respective $\sigma$-algebra. In agreement with this extension theorem, we can think that elements as the very fine partition of space, and $t_P$ as the very fine partition of time intervals, can be seen as a natural way of measuring the physical world, and do not necessarily means a discrete nature of space.

Now we turn our attention to the case of some signal moving with the speed of light $c$ along the positive $x$-axis. In this case case $y,z=0$, and $x=ct$. Further, we have $x'=ct'$. To see it, take the limit $u\to1$ in eqs.(\ref{inversask}),(\ref{inversask1}), $k=1$, consider eq.(\ref{gamma1}) for $\gamma_1$, and the third of eqs.(\ref{ecsyz1}) for $\beta_1$; one obtains $x'=ct'=\epsilon x/2$. This result enables us to examine the behaviour of eq.(\ref{luz_1}) in the scope of present measure theory of somas. We have that half of the structure associated to light is due to the soma $L$ (location), while the other half corresponds to the soma $T$ (time). A possible interpretation of this result is that physical measures associated to a signal propagating with the speed of light, are such that they are split in a half and half way, namely, the first part measured as $\epsilon^{-1}x'=x/2$ represents any element of space where the signal is at some moment of time, while the other half, that associated to measure $\epsilon^{-1}ct'$, indicates a tendency to move, to occupy an element of space along the direction of propagation, to be occupied a time $\epsilon^{-1}t'=x/(2c)$ later; the second element is located a distance $x/2$ ahead with respect to the former element. As both terms appear in eq.(\ref{luz_1}), then the two mentioned elements are integral part of the structure which is moving with the speed of light, becoming a continuous structure of elements, i.e. a collection of elements, of total extension $x=\epsilon^{-1}(x'+ct')$, in a half and half manner as previously discussed. Let $\lambda_1$ denotes the net extension of this structure moving with the speed of light, and let us use the Cartesian coordinates $x$,$y$,$z$ to describe the position of each of its elements; in vector form we can use eq.(\ref{vector-c_k}) for it, with $k=1$. At some instant, that signal propagating with the speed of light in vacuum along some direction given by the unit vector $\hat{{\bf e}}$ has a leading and a trailing element, say, elements $A_{le}$,$A_{te}$, respectively. Now, let ${\bf r}=c t\, \hat{{\bf e}}$ represents the position of the trailing element after some time $t$, assuming that it was at the reference element $A_o$ when a local clock there indicates $t=0$. Let us consider now that the position of any element of the structure moving with the speed of light in vacuum $c=c_1$ is given by a vector ${\bf r}$ with respect to the element $A_o$ when the clock indicates time $t$: ${\bf r}=(ct+\lambda_1\phi_1) \hat{{\bf e}}$, where $0\leq\phi_1\leq1$ is a number specifying the element of the structure; its null value ($\phi_1=0$) indicates the trailing element whereas the unit ($\phi_1=1$) indicates the leading element. Eqs.(\ref{vector-c_k}),(\ref{xt-luz-i}), enable us to extend this reasoning to any structure moving with speed $c_k$, for any $k\geq1$, in their linear approach. Thus, 
\begin{equation}\label{r-luzk}
{\bf r}=(c_kt+\lambda_k\phi_k) \hat{{\bf e}},
\end{equation}
\noindent is the position of any element of the structure which propagates with speed $c_k$, where the real number $0\leq\phi_k\leq1$ specifies the element. Let $\Phi_k=2\pi\phi_k$, and calls it the phase of the moving structure. Let us introduce the variables:
\begin{equation}\label{vec-K}
{\bf K}_k=\frac{2\pi}{\lambda_k} \hat{{\bf e}},\,\, \omega_k=\frac{2\pi}{\lambda_k}c_k.
\end{equation}
The magnitude of 3-vector ${\bf K}_k$/($2\pi$), represents the number of times structure's length fills the unit of length, when letting it to propagate freely along a direction given by the unit 3-vector $\hat{\bf e}$, and $2\pi /\omega_k=\tau_k$ denotes the time after which the structure repeats itself ahead at speed $c_k$; its inverse, $\nu_k=\tau_k^{-1}$, gives the number of times the structure has been completely displaced along the direction of propagation during the unit of time. One sees that $\omega_k=2\pi\nu_k$, and $\lambda_k\nu_k=c_k$. From eqs.(\ref{r-luzk}),(\ref{vec-K}) we have that:
\begin{equation}\label{phasek}
\Phi_k={\bf K}_k\cdot{\bf r}-\omega_k t.
\end{equation}
Due to the invariance of structure's speed $c_k$ for any inertial observer, eq.(\ref{phasek}) has the same value for inertial observers, if they are describing the same element; thus, if two inertial frames $\Sigma$,$\Sigma'$ are moving uniformly with respect to each other, we have that $\Phi_k(\Sigma)=\Phi_k(\Sigma')$. Taking into account this invariance of the phase, and using eqs.(\ref{phasek}),(\ref{ecsxtk}),(\ref{ecsyzk}), one obtains:
\begin{equation}\label{dopplerk}
\nu'_k=(1-lu_k)\gamma_k\nu_k,\,l'=\frac{l-u_k}{1-lu_k},\,m'=\frac{m\beta_k}{1-lu_k},\,n'=\frac{n\beta_k}{1-lu_k},
\end{equation}
where unprimed variables refer to variables described by inertial frame $\Sigma$ and the primed ones by $\Sigma'$, and $l$,$m$,$n$, stand for the direction cosines of 3-vector ${\bf K}_k$ with respect to frame $\Sigma$. Eqs.(\ref{dopplerk}) can be seen as a generalisation of Doppler's effect for the first of these equations, and the last three can be interpret as a generalisation of the aberration of light. If we replace eqs.(\ref{inversask})-(\ref{inversask2}) into eq.(\ref{phasek}), one obtains unprimed direction cosines similar in form as the expressions given by the three last of eqs.(\ref{dopplerk}), in which unprimed variables change to the respective primed ones and vice versa, and $u_k$ changes to $-u_k$; for instance, $l=(l'+u_k)/(1+l'u_k)$, etc. Also, one finds:
\begin{equation}\label{freq-lambda}
\nu_k=(1+l'u_k)\beta_k^{-2}\gamma_k^{-1}\nu'_k,\,\lambda_k=\frac{\beta_k^2\gamma_k}{1+l'u_k}\lambda'_k.
 \end{equation}
Going forward in our interpretation of the results presented here, we can think that when photons propagate, or particles move, what are moving are their associated structures defined on the elementary blocks of space, understanding by them the finest partition of space or elements as we call them. If at some instant $t$ the leading element ($\phi_k=1$) of a linear structure moving with speed $c_k$ in direction $\hat{{\bf e}}$ is located by ${\bf r_2}$, and if the trailing element ($\phi_k=0$) is positioned by ${\bf r_1}$, then according to eq.(\ref{r-luzk}) we have that ${\bf r_2}-{\bf r_1}=\lambda_k\hat{{\bf e}}$. To specify that this structure occupies an extension of space $\lambda_k$ at time $t$, in the direction $\hat{{\bf e}}$, we can introduce a function $F$ of a parameter $\xi={\bf r}\cdot\hat{{\bf e}}$, such that $F(\xi)=1$ if $\xi_1\leq\xi\leq\xi_2$, and $F(\xi)=0$ elsewhere.

\section{Dynamical measures}
We tackle now the problem of associating dynamical measures to every element of space of a structure moving with the speed of light in vacuum (say, a photon), in such a way that it becomes compatible with our previous interpretation. Nowadays, we know that photons have an energy $E_1=h\nu_1$, where $\nu_1$ is its associated frequency, and $h$ is Planck's constant. According to our interpretation, this structure covers an extension $\lambda_1$ in a half and half manner, in such a way that half of their elements are really "occupied", whilst the other half are ready to be occupied later. The complete situation replicates itself after a time $\tau_1=\lambda_1/c_1$, a distance $\lambda_1$ ahead in a given direction. Thus, we obtain for every occupied element of a structure moving with the speed of light, a dynamical measure $h=E_1\tau_1$. Space-time measures as given by eqs.(\ref{ecsxtk}),(\ref{ecsyzk}) applies to any element of space, assuming our interpretation. As the $\gamma_k$ measure factor increases with speed, we propose a generalisation of the previous consideration for light structures, by associating to every element of space, a main dynamical measure, called here the S-measure, given by:

\begin{equation}\label{S-measure}
S(u)=\gamma_kS_o=E_k\tau_k,\, \text{if}\,\, u\neq 0,\quad \text{and:}\,\, S(0)=(1+\epsilon^2)^{-1/2}S_o,
\end{equation}

\noindent where $S_o=\epsilon h$ is a constant. Thus, $S(1)=h$, $S(c_2/c)=2h$, etc. For the case $S(0)$ there is not an energy-time product, because there is not a propagation. Let us consider a structure moving with the speed of light, with a number of elements $N=\lambda_1/L_P$  associated to the structure, half of them (i.e. $N/2$) have at some moment $S=h$, while the other half of the elements have $S\simeq S_o$ in vacuum. Similarly, we have that S-measures associated to occupied elements of structures with speed $c_k$, are equal to $S(c_k/c)=kh=E_k\tau_k$; in these cases $\tau_k=\lambda_k/c_k$, thence: $E_k=khc_k/\lambda_k$. Thus, we have a suitable way to deal with dynamical properties of different kind of structures defined on space-time in terms of elements. Let us define now the dimensionless speeds $U^*$, whose squared values depend on the regime of motion as follows: 

\begin{align}
U^{*2}= &\, 1-3\epsilon^2,\,\text{if}\, k=1, \label{particle1}\\
U^{*2}= &\, (c_k/c)^2-3\epsilon^{2(k-1)},\,\text{if}\, k\geq2.\label{particlek}
\end{align}

Thence, the S-measure $S(U^*)$ equals $(2k-1)h/2$, that is, $h/2$, $3h/2$, $5h/2$, etc., for $k=1,2,3$, and so on, respectively. Ponderable matter particles like electrons and protons can be accelerated from the rest up to relativistic speeds $v<c$. As $U^*<1$ in the subluminal regime of motion, one can have a structure which as a whole is at rest with respect to some inertial frame of reference, though their occupied elements have a S-measure $h/2$ each. Let us assume a structure with a length at rest $\lambda_o$, in an almost half and half situation as considered previously, that is, half of structure's elements are occupied and almost the other half will be occupied later after a time $\tau_o=\lambda_o/(2cU^*)$ in a back and forth mode, with $U^*$ given by eq.(\ref{particle1}). This structure can also be in motion, say, with velocity $\bf{v}$. In this case, if the occupied elements have $S=h/2$ each, then we have that $\mathbf{U}^*=\mathbf{U}^{\dagger}\oplus\mathbf{u}$, where $\mathbf{u}=\mathbf{v}/c$ and $\mathbf{U^{\dagger}}$ represents some kind of internal motion and the operation $\oplus$ stands for the relativistic sum of velocities -cf. eqs.(\ref{sumaE}),(\ref{speedsk}). In this way we can consider two different cases for this kind of structure, say, one in which it is at rest and the other where it is moving, both with respect to some inertial frame. We have per occupied element: 

\begin{equation}\label{E1}
E_1\tau_1=E_o\tau_o=h/2.
\end{equation}

\noindent where $E_o$ is particle's rest energy and $\tau_o=\lambda_o/(2U^*c)$ is the associated internal structure time at rest. Therefore, for the case of the structure at rest, we obtain:
\begin{equation}\label{Compton}
\lambda_o=\frac{hc}{E_o}U^*\simeq\lambda_C,
\end{equation}
\noindent where $\lambda_C$ is particle's Compton wavelength. We made above the approximation $U^*\simeq1$. From eqs.(\ref{S-measure}),(\ref{E1}) we find the expression for the energy associated to every element of a structure in the subluminal regime of motion, taking (i) $\tau_1=\lambda_1/(U^{\dagger}c)$, $\tau_o=\lambda_o/(U^*c)$, $\lambda_o\simeq\lambda_C$  and (ii) considering that $\lambda_k/\lambda_o$ behaves as the length of a body in motion and consequently is given by the first of eqs.(\ref{lengthsk}). We use here $\tau_o$ without the factor $1/2$ which appeared when we derived eq.(\ref{Compton}), for in that case we considered a structure in a back and forth mode, and present case it is not so, though structure's rest length $\lambda_o$ is the same. Further, we shall consider $U^{\dagger}$ in eq.(\ref{speeds-inv}) in the role of $U'_1$, while for $U_1$ there we take $U^*$ given by eq.(\ref{particle1}). With previous arguments we find:
\begin{equation}\label{Energy}
E_1=\frac{E_o U^{\dagger}}{\beta_1^2\gamma_1 U^*}\simeq\frac{E_o}{\sqrt{1-v^2/c^2}},
\end{equation}

\noindent where the approximation above holds when $\beta_1^2\gg\epsilon^2$, that is, for relativistic speeds. We can not apply eq.(\ref{Energy}) for the case $u=1$, i.e. for the speed of light in vacuum ($v=c$). As we shall discuss later, if a ponderable matter particle transits from the subluminal regime to the first superluminal regime, or in the converse sense, it should be done in a discrete way, such that changes in $u^2$ are expressed in terms of $\epsilon^2$ around the unit. For structures whose occupied elements have associated a speed $U^*$ and rest length $\lambda_o\simeq\lambda_C$, the S-measure of such elements change from $h/2$ each to $3h/2$ in the $k:1\to2$ transition. Let us consider the energy associated to the occupied elements of these structures in any superluminal regime of motion, $k\geq2$. From eqs.(\ref{S-measure}),(\ref{particlek}) we can write:

\begin{equation}\label{Energyk1}
E_k=(2k-1)\frac{E_o\tau_o}{\tau_k},\,\tau_k=\frac{\lambda_k}{c U^{\dagger}},
\end{equation}

\noindent where we used $E_o\tau_o=h/2$. For $\tau_o$ we consider the same expression as used previously, and for $\tau_k$ we take into account eq.(\ref{particlek}) to calculate $U^{\dagger}_k$, and thereafter we take $U^{\dagger}$=$(c_k/c)U^{\dagger}_k$. Using eqs.(\ref{speeds-inv}),(\ref{particlek}) we arrive to:

\begin{equation}\label{interno}
U^{\dagger2}_k=1-\frac{3\epsilon^{2(k-1)}(c/c_k)^2(1-u^2_k)}{(1-{\bf u_{k}}\cdot{\bf U^*_{k}})^2}.
\end{equation}

As $c_k/c\simeq\epsilon^{-k+1}$ for all $k\geq2$, then in a first approximation we can consider $U^{\dagger2}_k\simeq1$ and then $U^{\dagger}\simeq\epsilon^{-k+1}$. Further, we take $\lambda_k/\lambda_o=\beta^2_k\gamma_k$, and approximates it to $(k-1)(1-u^2_k)\epsilon^{-1}$, valid if $(1-u_k^2)\gg\epsilon^{4(k-1)}$, a reasonable assumption. From these considerations and eq.(\ref{Energyk1}) we find:
\begin{equation}\label{Energyk2}
E_k\simeq\frac{2k-1}{k-1}\frac{\epsilon^{-k+2}E_o}{1-u^2_k},\, k\geq2.
\end{equation}
Thus, for the first superluminal regime we have: $E_2\simeq3E_o/(1-u^2_k)\simeq3E_o$, where the last approximation is valid even for very large speeds, say, for values of order $u\sim10^8$. This result is useful, for one can accelerate a particle, for instance an electron, from the rest up to an energy $3E_o$, $u^2=8/9$, and apply a S-measure $h$ to all occupied elements of it and almost $h$ (a bit less) for the other elements, then it makes the discrete transition from the subluminal regime to the first superluminal regime of motion. The non occupied elements of the structure have $S\simeq3S_o$ each in the subluminal regime, and after the transition each one has a S-measure $S\gtrsim h$. In this transition, particle's energy almost remains the same.

We want to explore now the S-measure associated to elements of vacuum between charged particles in electrostatic interaction along field lines. Let $U_e$ be defined as:
\begin{align}
U_e^2= &\, 1-N_e\epsilon^2,\,\,\text{if}\,\, k=1, \label{electro1}\\
U_e^2= &\, (c_k/c)^2-N_e\epsilon^{2(k-1)},\,\text{if}\,\, k\geq2,\label{electrok}
\end{align}
such that:
\begin{equation}\label{Se}
S(U_e)=\left(k-1+\frac{\alpha}{2\pi}\right)h,\quad \text{for all}\,\,k\geq1,
\end{equation}
where $\alpha$ is fine's structure constant. The number $N_e$ given in eqs.(\ref{electro1}),(\ref{electrok}) equals:
\begin{equation}\label{Ne}
N_e=\left(\frac{2\pi}{\alpha}\right)^2-1\simeq7.41\times10^5.
\end{equation}
We used the form $E_k\tau_k$ given in eq.(\ref{S-measure}) to infer the $S(U_e)$ value, $k=1$, of eq.(\ref{Se}). In effect, let us consider an electron interacting with another electron. The Coulomb potential energy associated to this interaction is given by $e^2/(4\pi\varepsilon_0 r)$, where $r\gg\lambda_C$ is the separation between the charges and $e$ is the elementary charge; the interaction time is of order $\tau_1=r/c$, then its product equals $\alpha\hbar$. The $U_e$ given by eqs.(\ref{electro1}),(\ref{electrok}) can be seen as an open part of the structure at the end element of it such that $\bf{U}_e$ is parallel to the linear structure representing the particle. That is, the structure at rest with length $\lambda_o$ is moving along in a back and forth manner, but at the end there is a flow which extends off the structure, and therefore affects the space around it. If we have two conductor plates where in one of them are electrons in excess and in the other there are an equal lack of them, we have electric field lines in  the region between the plates. In that situation, we have that the open part of the structure associated to each electron spreads out in various field lines in the space between the plates, such that the elements of them have a dynamical S-measure $S_f=S(U_f)=S(U_e)/N_f$, where $N_f$ is the number of field lines. There should be a limit to $N_f$, say, $N_{max}$, because from eq.(\ref{S-measure}), $k=1,$ one expects that when $U_f\to\epsilon$ then $S_f\to S_o$, and $u^2=\epsilon^2$ is the previous value before the null value $u^2=0$ in the subluminal regime of motion. From eq.(\ref{Se}) and considering that $S_o=\epsilon h$, we write the limit condition as:
\begin{equation}\label{Nmax}
\epsilon\,N_{max}=\frac{\alpha}{2\pi}.
\end{equation}

Thus, if we infer a value for $N_{max}$ we can find an estimated value of $\epsilon^2$, which is a fundamental constant in this work. We can consider that when $N_f\to N_{max}$, it is like a dilution of field lines into vacuum until $S_f$ reach the limit $S_o$. The number of elements associated to a structure of ponderable matter particle like an electron is of the order of $\lambda_o/L_P\simeq\lambda_C/L_P\sim10^{23}$. Let $N_o=N_A\times(1\,\text{mol})$, where $N_A$ is Avogadro's number, that is, $N_o$ is a dimensionless number. We consider here that $N_{max}\simeq N_o$ is a good choice on the following bases. As first calculated by Schwinger \cite{Schwinger} electron magnetic moment anomaly in first approximation equals $a_e\simeq\alpha/(2\pi)$. The number of elements participating on measurements affecting an electron are of order of $\lambda_o/L_P$. Present day best measurements of both the electron magnetic moment anomaly $a_e$ \cite{CODATA2014} and of many isotope masses \cite{masas2013} come from experiments carried out using Penning traps \cite{geonium1986}. These traps are built using a uniform magnetic field superposed to a quadrupole electric field. The former field implies a cyclotron frequency inversely proportional to the masses of involved charged particles. Actual standard of atomic masses takes one mole of pure Carbon-12 atoms as exactly equal to 12 grams of mass. Thus, if we take $N_{max}=N_o$ and replace it into eq.(\ref{Nmax}), we obtain an estimated value for our fundamental constant $\epsilon^2$ as: 

\begin{equation}\label{epsilon2}
\epsilon^2 = \left(N_o^{-1}\frac{\alpha}{2\pi}\right)^2 \simeq 3.72\times10^{-54},
\end{equation}
\noindent which is an extremely tiny constant. From it, we see that $\epsilon\simeq1.9\times10^{-27}$, $\epsilon^{-1}\simeq5.2\times10^{26}$, and $S_o=\epsilon h\simeq1.3\times10^{-60}$ Js. Previously, we saw that $c_k/c\simeq\epsilon^{-k+1}$, $k\geq2$, then $c_{k+1}/c_k\simeq\epsilon^{-1}\gg1$.

\section{Heuristic experiment}

We propose here an experiment to energise/de-energise photons, both as a test of our interpretation of the non divergence of our space-time $x,t$ measures as given by eqs.(\ref{Leps-1}),(\ref{ecsxt1}) for $u=1$ or $v=c$, as well as for the form which acquire these expressions, given by eq.(\ref{luz_1}). According to our theoretical construction, let us consider the space between two electrically polarised plates with field lines going from the negative to the positive plate, for each free electron attached to the negatively polarised plate spreads out its open $S(U_e)$ into many field lines, say $N_f$, as considered previously, such that $S(U_e)=N_fS_f$. The S-measure $S_f$ satisfies eq.(\ref{S-measure}), for $k=1$; in this case we have $\beta_1\gg\epsilon^2$ (remember that $N_e\gg1$). Thus:
\begin{equation}\label{Sf}
S_f=S(u_E)\simeq S_o(1-u^2_E)^{-1/2},
\end{equation}
where $u_E$ is the dimensionless speed of the elements of each field line. If a low speed electron enters this region, its motion is altered until it follows a field line and does it with speed $u_E$, so after that its kinetic energy $E_{kin}$ approximately equals:
\begin{equation}\label{Ekin}
E_{kin}\simeq\left[(1-u_E^2)^{-1/2}-1\right]E_o,
\end{equation}
where $E_o\simeq511$keV is electron rest energy. Let us consider now a photon with wavelength $\lambda_{in}$ which comes into the field region, going from the negative to the positive polarised plate. If it goes along a field line and after some time emerges from it, one sees it from the laboratory frame as a photon emitted by an inertial reference frame moving with respect to the laboratory with some dimensionless speed $u_E$, which can be calculated using eq.(\ref{Ekin}). In effect, if the electric potential difference applied between the plates is $V$ volts, we can write down in eq.(\ref{Ekin}) $E_{kin}=eV$. The second of eqs.(\ref{freq-lambda}) gives us the ratio $\lambda_{out}/\lambda_{in}=\eta$, where $\lambda_{out}$ is the wavelength of the emerged photon. In that equation one takes $\lambda'_1$, $\lambda_1$ as $\lambda_{in}$, $\lambda_{out}$, respectively. Further, we take there $l'=1$, $\beta_1^2\gamma_1\simeq\beta_1$, then one obtains:
\begin{equation}
\eta^2=\left(\lambda_{out}/\lambda_{in}\right)^2\simeq\frac{1-u_E}{1+u_E},\quad\text{if}\,\,\eta<1.
\end{equation}
From this equation we solve for $u_E$, replace into eq.(\ref{Ekin}), and arrive to:
\begin{equation}\label{voltage}
\frac{eV}{E_o}=\left[1-\left(\frac{1-\eta^2}{1+\eta^2}\right)^2\right]^{-1/2}-1.
\end{equation}
From this expression we see that if $\eta$ takes the values 1/2, 1/3, 1/4, 1/5, 1/10, just to mention some examples, we find for $eV/E_o$: 1/4, 2/3, 9/8, 8/5, 81/20, respectively, or in terms of electric potential difference we have the values 127.75kV, 340.67kV, 574.87kV, 817.6kV, and 2.07MV, correspondingly. These values are attainable in laboratory, at least the first three of them, then there exists a real chance to test present interpretation for structures moving with the speed of light in vacuum. In the previous considered values for $\eta$ we have the case of a process of energisation of photons. It is necessary to warrant that the photon goes along a field line, this is the key point in the proposed experiment. If experiments confirm present heuristic reasoning, we have that a photon comes in and a photon comes out with different energy. In the case $\eta=1/2$, nonlinear optics \cite{Boyd2008} uses special crystals to carry out it, but the difference with respect to our proposed 1-1 photon procedure lies in that they need two photons of the same wavelength, say $\lambda_{in}$, from a laser source to obtain a single photon of wavelength $\lambda_{out}=\lambda_{in}/2$. Further, our $\eta$ can be a non rational number. If we reverse the polarity between the plates maintaining the direction of photon propagation the result is a de-energisation, which changes $\eta$ by its inverse in eq.(\ref{voltage}). If we design an experiment to verify our hypothesis of photon's energisation by electric fields, maybe we want to test $\eta$ as a function of the applied voltage. In that case, instead of eq.(\ref{voltage}) we can consider the following expression:

\begin{equation}\label{eta}
\eta=\left\{\dfrac{1\mp[1-(\frac{eV}{E_o}+1)^{-2}]^{1/2}}{1\pm[1-(\frac{eV}{E_o}+1)^{-2}]^{1/2}}\right\}^{1/2},
\end{equation}

\noindent where the upper sign is taken for photons going from the negative to the positive polarity and the lower sign for the converse direction. If we take a beam of photons, maybe the superposition of an axial magnetic field with the same polarity of the applied electric field could increase the efficiency of energisation. If our heuristic reasoning leading us to an energisation of photons in a 1-1 process is confirmed by experiments, it  could be useful in fusion research. For instance, Atzeni and Meyer-ter-Vehn in their book on Inertial Fusion \cite{Fusion}, table 3.1, reported the main results of a simulation for the implosion of a deuterium-tritium pellet due to the impact of radially incoming photons from a powerful laser source, looking for the nuclear fusion of the hydrogen isotopes. They assumed photons of 0.25$\mu$m of wavelength and a total light energy of 1.7MJ; the reported energy gain is 65, a very good figure. If we have a laser source of 1.06$\mu$m photons and total energy of order of 1MJ, then if our proposed experiment is confirmed by experiment, we can think of taking these photons and energise them by a factor of 4.24, or $\eta=0.25/1.06$, thence obtaining a good scenario for the wanted fusion, assuming a photon energisation efficiency of 0.5.

\section{Discussion}
\begin{figure}
\centering
\includegraphics[scale=0.4]{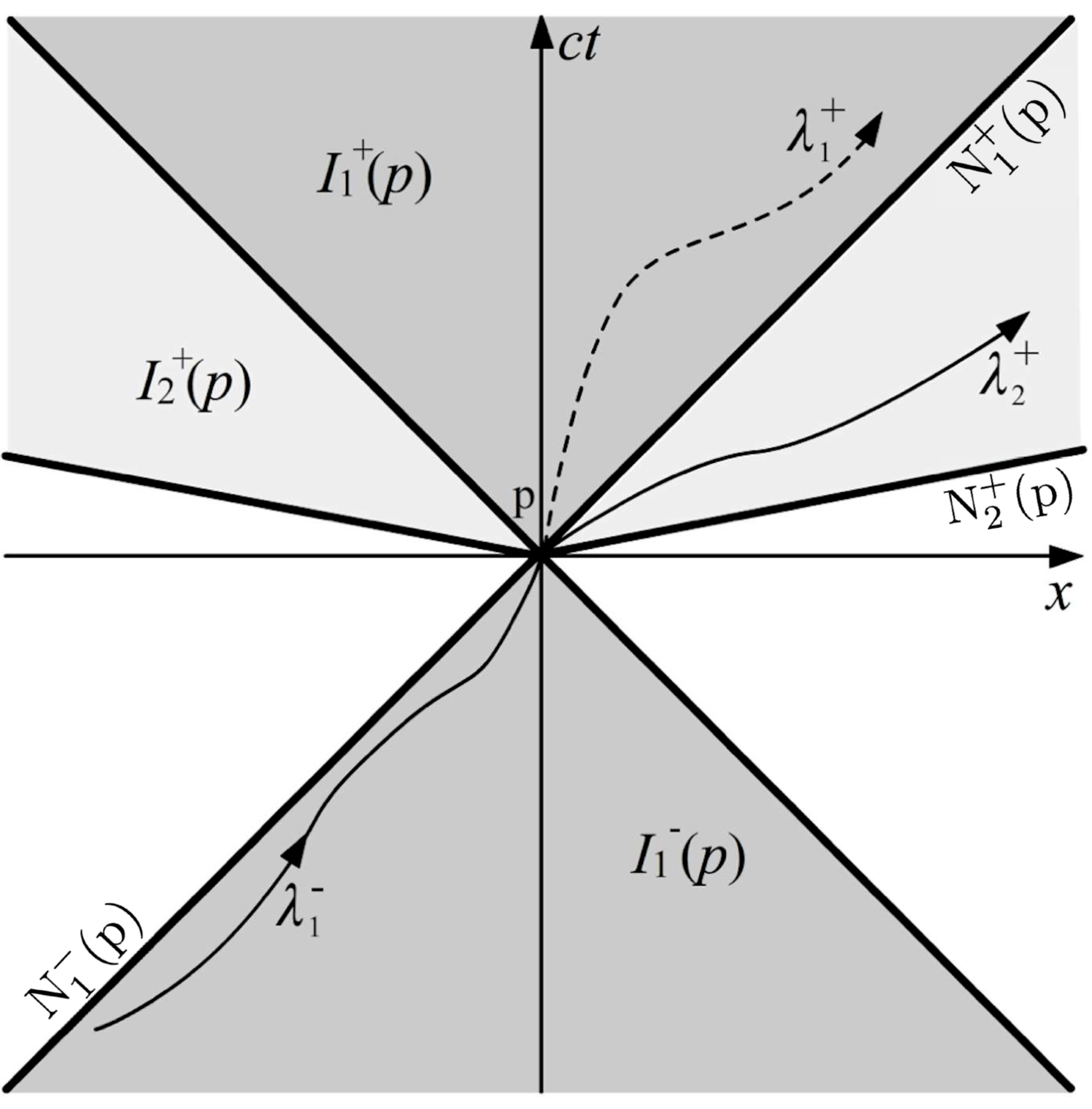}
\caption{A ponderable matter particle could make a transition from the subluminal regime of motion to the first superluminal regime at some event $p$, through a discrete jump in $u^2=v^2/c^2$ around the unit in terms of $\epsilon^2$.}\label{transition1-2}
\end{figure}

In present work we extended the notion of timelike curves, such that we have k-timelike curves, depending on the regime of motion in which we describe material particles.  Thus, the usual timelike curve, that is, in the subluminal regime of motion, $k=1$, is now an 1-timelike curve. Let us consider the theoretical possibility of a change of the regime of motion for material particles. Such change can not be done in a continuous manner, because if a material particle is accelerated starting from the rest till it reaches some speed near but less than the speed of light in vacuum (say, $v\simeq0.94c)$, all this process is described in space-time by a 1-timelike path, and if it surpasses the speed of light it should be described by a 2-timelike curve; during this transition it could not move with the speed of light, because if it does, ceases to be a ponderable matter particle. Then, the transition from the subluminal regime to the first superluminal regime of motion ($k=2$) around the speed of light in vacuum, that is, around the $u^2=1$ value, should be done in a discrete way; as both, the $\gamma_1$ and $\gamma_2$ factors contain the constant $\epsilon^2$ as the finest quantity there, then we have that changes in $u^2$ are quantified in terms of $\epsilon^2$:
\begin{equation}\label{cambio_u2}
u^2: 1-n_1\epsilon^2\rightleftarrows1+n_2\epsilon^2,
\end{equation}
\noindent where $n_1$,$n_2$ are positive real numbers, and the cited transition is given by the left to right arrow above. The other direction, the right to left transition, corresponds to a change from the first superluminal regime of motion to the subluminal one. For charged particles or material bodies we should consider $n_1,n_2\gg1$, for they must be appreciable larger than the $N_e$ given by eq.(\ref{Ne}), in order that the electronic structure is not altered by the transition. According to  eq.(\ref{cambio_u2}) net changes in $u^2$ are $\pm(n_1+n_2)\epsilon^2$ where the plus sign applies for the subluminal to superluminal transition. We illustrate the transition from the subluminal to the first superluminal regime of motion in Fig.\ref{transition1-2}; there, $p$ is some event of space-time where a transition happens as described by eq.(\ref{cambio_u2}). Before the change, the particle followed the timelike path $\lambda_1^{-}$, and after the transition it moves following the timelike path $\lambda_2^{+}$.

One of the goals of NASA's Breakthrough Propulsion Physics Program (BPPP), as reported by Millis \cite{Millis}, was looking for a possible transition around the speed of light value for a material body. The answer to this question was negative in the report, for present day physics does not allow it due to the continuous character of speed changes, which implies an infinite energy consumption. Present work enables to answer that BPPP's question affirmatively, as inferred from our considerations above and eq.(\ref{cambio_u2}). There is a proposition of "warp drive" made by Alcubierre \cite{Alcubierre}, which uses a special kind of matter which has the time component of the energy-momentum tensor negative ($T^{00}<0$). We propose instead, to take a material particle or body, which can start from the rest, accelerates it till certain speed near but less than the speed of light in vacuum, following an 1-timelike curve, and at some event (say, $p$) make a discrete transition in $u^2$ according to eq.(\ref{cambio_u2}), then it changes its path in space-time to a 2-timelike curve. In this way, the incoming 1-timelike curve of the particle, say, $\lambda_1^{-}$, belongs to the region $I_1^-(p)$ and the path after the discrete change in $u^2$, say, $\lambda_2^{+}$ lies in the region $I_2^+(p)$. In Fig.\ref{transition1-2} the dotted path in region $I_1^+(p)$ would be the path followed by a particle if no discrete change in $u^2$ occurs at the event $p$. To carry this change out, we should take into account the associated S-measure changes, i.e. we should add a $\Delta S=h$ per occupied element. For the non occupied elements of the particle, the change in S-measure, $\Delta S$, should be calculated using eqs.(\ref{S-measure}),(\ref{gamma2}), given approximately by:

\begin{equation}
\Delta S\simeq\{1+\epsilon^2(1-u_2^2)^{-1/2}\}h-S_i,
\end{equation}

\noindent where $S_i$ is the initial S-measure of these elements before the discrete change in $u^2$. For instance, if the initial $u^2=8/9$, $S_i\simeq3S_o$, then, we have that this $\Delta S$ is only a bit less than h.
\section{Conclusions}
In present article we explored the consequences of a generalisation of the local causality principle of space-time. The generalisation is considered in the sense that local causality is valid even for material particles traveling with speeds greater than the speed of light in vacuum $v>c$. We proposed the existence of a set of metrical null cone speeds $\{c_1, c_2, c_3, \ldots\}$, where $c_1=c$. In this way we partioned dimensionless speeds $v/c$ in intervals, where only the first one is closed, called here the subluminal regime of motion, $k=1$, while the other intervals are half-closed at the right, that is, the upper value is $c_k/c$ for $k\geq2$ and we call these intervals superluminal regimes of motion, where we denoted the interval or regime of motion by a natural index $k$. For all regimes of motion we found space-time measures which (1) do not diverge at the maximum speed of the regime of motion $c_k$, and which (2) implies a generalised form of Einstein's addition of velocities. The non divergent space-time measures in the subluminal regime reduce to Lorentz transformations in first approximation. The non divergence for $v=c$ of these equations is reached by adding a very tiny constant, $\epsilon^2$, into the Lorentz radical appearing in the $x$ and $c_1t$ equations, and we denoted this new factor as $\gamma_1$, which equals $\epsilon^{-1}$ when $v=c$. To keep invariant the speed of light in vacuum for inertial observers under uniform relative motion, we modified the expressions for $y$ and $z$ measures, too, such that $y/y'$, and $z/z'$ equal $\beta_1\gamma_1$, where $\beta_1=(1-v^2/c_1^2)^{1/2}$. We generalised this set of space-time measures replacing previous subindex 1 by an index $k\geq1$, and obtained eqs.(\ref{ecsxtk}),(\ref{ecsyzk}). We imposed the conditions that $\gamma_k=k\epsilon^{-1}$, $\beta_k=0$, for $v=c_k$. In matrix form space-time measures are given by eqs.(\ref{Lk}), valid for all regimes of motion, which are interpred as a generalised Lorentz transformation followed by a regularisation factor $\beta_k\gamma_k$. From the matrix expression we found the inverse measures.

For arbitrary $k\geq2$, that is, for all superluminal regimes, we found $\gamma_k$, $c_k$, which are given by eqs.(\ref{gammai}),(\ref{luz-i}), respectively. Metrical null cone speeds ($k\geq2$) approximately equals $c_k\simeq\epsilon^{-k+1}c$. We found also expressions for time intervals and lengths of bodies in motion.

We introduced a 2-form, a k-interval and a k-metric given by eqs.(\ref{2-form}),(\ref{k-interval}),(\ref{k-metric}), respectively. With these constructions we talk of k-timelike and k-null intervals, and of k-timelike and k-null paths which can be traced from/to a given event. Thus, we can connect event pairs through these kind of paths and obtain a partial ordering on space-time, such that we can speak that some event $p$ causally or chronologically precedes other event $q$ for all regimes of motion. With these paths we constructed a causal structure on space-time. Future and past regions of space-time causally or chronologically related to some given event $p$, are denoted here by $J_k^{\pm}(p)$, $I_k^{\pm}(p)$, respectively, where the plus sign denotes future zones whilst the negative sign stands for past regions.

As space-time measures for a signal moving with the speed of light in vacuum is non divergent we interpreted this result. These expressions reduce the description to only one degree of freedom, along the direction of propagation -cf. eq.(\ref{luz_1}), because $y=z=0$, and $x=ct$; further, $x'=ct'$. We considered then the finest partition of space as of order of $L_P^3$, the cube of Planck's length and called it an element of space, and for time intervals the minimum partition is Planck's time $L_P/c$. We used Carath\'eodory's measure theory as a suitable mathematical ground to interpret our space-time measures. As a next step, the structure made of elements moving with the speed of light in vacuum has some length $\lambda$, of which half of its elements are occupied and the other half will be occupied a time $\lambda/(2c)$ latter, and the situation replicates completely after a lapse of time $\lambda/c$. This interpretation led us to construct a phase and from it we derived both a generalisation of Doppler effect and of the aberration of light. For the subluminal regime of motion, the expression for aberration coincides with that of relativity theory, while the one for Doppler's effect reduce to the one in relativity in first approximation, when $\beta_1\gg\epsilon^2$. To complement this interpretation, we introduced a simple dynamical measure called here S-measure -cf.eq.(\ref{S-measure}), which equals $h$ for every occupied element of a photon, and $h/2$ for occupied element of electrons in the subluminal regime of motion. We derived energy expressions for subluminal and superluminal motions for material particles. Curiously, the energy associated to the occupied elements equals three times the rest energy for both, the subluminal motion at $u^2=8/9$, and when it is in the first superluminal regime of motion with $v/c_2$ inappreciable compared with the unit. Thus, we can try to do the subluminal to superluminal transition -cf. eq.(\ref{cambio_u2})- with electrons whose speeds are of order $v\simeq0,94c$. It needs the addition of an amount of S-measure a bit less than $N_oh$ per electron, in the exact region where it lies. Using S-measures associated to charged particles like an electron we inferred a value for our constant $\epsilon^2\simeq3.7\times10^{-54}$. An experiment to energise photons is proposed as a test of our interpretation. The experiment considers a photon of some wavelength $\lambda_{in}$ following an electric field line, from the negative to the positive polarity, with a potential difference of order 127.75kV (kilo Volts), and after leaving the path, the photon comes out with $\lambda_{out}=\lambda_{in}/2$, in a one to one photon process. Eq.(\ref{voltage}) gives various possible outcomes for the photon's wavelength conversion, not necessarily $\lambda_{out}/\lambda_{in}$ a rational number, though various values as examples are given, also in a one to one photon energisation. For instance, if the energisation factor is 3, the required potential difference is of order of 340.67kV. This proposed experiment, if successful, can be applied to the search of inertial confinement fusion, for it involves the implosion of a pellet containing the nuclear "fuel" material, and the wavelength of the photons driving the implosion is of importance in this process.

Finally, we conclude that the generalisation of the local causality of space-time leaded us to new space-time measures, and consequently to a causal structure built on space-time by regions associated to regimes of motion, which besides with our interpretation in terms of elements of space, and the S-measure as the fundamental dynamical measure per element, allow us to think of making a discrete transition between the subluminal and the first superluminal regime of motion for material particles, as discussed in present work.

\section{Acknowledgements}
The author acknowleges the kind assistance of J.C. Buitrago-Casas (UC Berkeley) and J.S. Castellanos Dur\'an (Max Planck Solar) with some LaTeX issues. JCBC also put in digital form the original hand made drawings of the two figures. The author expresses his thanks to Dr. J.C. Mart\'{\i}nez-Oliveros (SSL, UC Berkeley) for encouraging him to write this paper.%



\bibliographystyle{RS} 
\bibliography{ModelingBidders}

\begin{thebibliography}{99}

\bibitem{Einstein1916}
{Einstein} A. 1916  {Die Grundlage der allgemeinen Relativit{\"a}tstheorie}.
  {\em Annalen der Physik} \textbf{354}, 769--822.

\bibitem{Hawking-Ellis}
{Hawking} SW, {Ellis} GFR. 1973 {\em {The large-scale structure of
  space-time.}}
Cambridge Univ. Press.

\bibitem{Hill-Cox2012}
{Hill} JM, {Cox} BJ. 2012  {Einstein's special relativity beyond the speed of
  light}. {\em Royal Society of London Proceedings Series A} \textbf{468},
  4174--4192.

\bibitem{Einstein1905}
{Einstein} A. 1905  {Zur Elektrodynamik bewegter K{\"o}rper}. {\em Ann. der
  Phys.} \textbf{322}, 891--921.

\bibitem{Minkowski}
{Minkowski} H. 1909  {Raum und Zeit}. {\em Physikalische Zeitschrift}
  \textbf{10}, 104--111.

\bibitem{Caratheodory}
{Carath\'eodory} C. 2010 {\em {Algebraic theory of measure and integration}}.
American Mathematical Society.

\bibitem{Zermelo1904}
{Zermelo} E. 1904  Beweis, dass jede Menge wohlgeordnet werden kann. {\em
  Matematische Annalen} \textbf{59}, 514--516.

\bibitem{Zeeman1964}
{Zeeman} EC. 1964  {Causality Implies the Lorentz Group}. {\em Journal of
  Mathematical Physics} \textbf{5}, 490--493.

\bibitem{Kronheimer-Penrose1967}
{Kronheimer} EH, {Penrose} R. 1967  {On the structure of causal spaces}. {\em
  Proceedings of the Cambridge Philosophical Society} \textbf{63}, 481.

\bibitem{Carter1971}
{Carter} B. 1971  {Causal structure in space-time}. {\em General Relativity and
  Gravitation} \textbf{1}, 349--391.

\bibitem{Geroch1970}
{Geroch} R. 1970  {Domain of Dependence}. {\em Journal of Mathematical Physics}
  \textbf{11}, 437--449.

\bibitem{Planck1899}
{Planck} M. 1899  {\"Uber irreversible Strahlungs-vorg\"ange, 5 (Schluss)}.
  {\em Sitzungsber. Preuss. K\"onigl. Akad. Wissen. zu Berlin} \textbf{I},
  440--480.

\bibitem{Sakharov1967}
{Sakharov} AD. 1967  {Vacuum quantum fluctuations in curved space and the
  theory of gravitation}. {\em Akademiia Nauk SSSR Doklady} \textbf{177}, 70.

\bibitem{Gravitation}
{Misner} CW, {Thorne} KS, {Wheeler} JA. 1973 {\em {Gravitation}}.
W.H. Freeman and Co.

\bibitem{Smolin2001}
{Smolin} L. 2001 {\em {Three roads to quantum gravity}}.
Basic Books.

\bibitem{Schwinger}
{Schwinger} J. 1948  {On Quantum-Electrodynamics and the Magnetic Moment of the
  Electron}. {\em Physical Review} \textbf{73}, 416--417.

\bibitem{CODATA2014}
Mohr PJ, Newell DB, Taylor BN. 2016  CODATA recommended values of the
  fundamental physical constants: 2014. {\em Rev. Mod. Phys.} \textbf{88},
  035009.

\bibitem{masas2013}
{Myers} EG. 2013  The most precise atomic mass measurements in Penning traps.
  {\em International Journal of Mass Spectrometry} \textbf{349}, 107--122.

\bibitem{geonium1986}
Brown LS, Gabrielse G. 1986  Geonium theory: Physics of a single electron or
  ion in a Penning trap. {\em Rev. Mod. Phys.} \textbf{58}, 233--311.

\bibitem{Boyd2008}
{Boyd} RW. 2008 {\em {Nonlinear Optics}}.
Academic Press.

\bibitem{Fusion}
{Atzeni} S, {Meyer-ter-Vehn} J. 2004 {\em {Inertial Fusion}}.
Oxford University Press.

\bibitem{Millis}
{Millis} MG. 2005  {Assessing Potential Propulsion Breakthroughs}. {\em Annals
  of the New York Academy of Sciences} \textbf{1065}, 441--461.

\bibitem{Alcubierre}
{Alcubierre} M. 1994  The warp drive: hyper-fast travel within general
  relativity. {\em Classical and Quantum Gravity} \textbf{11}, L73--L77.

\end{thebibliography}

\end{document}